\documentclass[12pt,preprint]{aastex} 

\usepackage{graphicx}
\usepackage[twocolumn]{emulateapj5}

\shortauthors{SCHMIDT ET AL.}
\shorttitle{THE ECLIPSING POLAR SDSSJ~015543.40+002807.2}

\received{}
\revised{}
\slugcomment{To appear in Astrophysical Journal}

\begin{document}

\tolerance=10000

\title{Unravelling the Puzzle of the Eclipsing Polar SDSSJ~015543.40+002807.2 with
{\it XMM} and Optical Photometry/Spectropolarimetry\altaffilmark{1}}

\author{
Gary D. Schmidt\altaffilmark{2},
Paula Szkody\altaffilmark{3},
Lee Homer\altaffilmark{3},
Paul S. Smith\altaffilmark{2},
Bing Chen\altaffilmark{4},
Arne Henden\altaffilmark{5},
Jan-Erik Solheim\altaffilmark{6},
Michael A. Wolfe\altaffilmark{3},
and
Robert Greimel\altaffilmark{7}
}

\altaffiltext{1}{Some of the results presented here were obtained with the MMT
Observatory, a facility operated jointly by The University of Arizona and the
Smithsonian Institution.  Additional results are based on observations made
with the Nordic Optical Telescope, operated on the island of La Palma jointly
by Denmark, Finland, Iceland, Norway, and Sweden, in the Spanish Observatorio
del Roque de los Muchachos of the Instituto de Astrofisica de Canarias.}

\altaffiltext{2}{Steward Observatory, The University of Arizona, Tucson AZ
85721; gschmidt@as.arizona.edu, psmith@as.arizona.edu}

\altaffiltext{3}{Department of Astronomy, University of Washington, Box 351580,
Seattle, WA 98195-1580; szkody@astro.washington.edu,
homer@astro.washington.edu, maw2323@u.washington.edu}

\altaffiltext{4}{XMM-Newton Science Operations Centre, ESA/Vilspa, 28080,
Madrid, Spain, bchen@xmm.vilspa.esa.es}

\altaffiltext{5}{Universities Space Research Association/US Naval Observatory,
Flagstaff Station, P.O. Box 1149, Flagstaff, AZ 86002-1149; aah@nofs.navy.mil}

\altaffiltext{6}{Institute of Theoretical Astrophysics, P.O. Box 1029,
Blindern, Oslo, N-0315, Norway, j.e.solheim@astro.uio.no}

\altaffiltext{7}{ING, Apartado de Correos, 321, 38700 Santa Cruz de La Palma,
Canary Islands, Spain, greimel@ing.iac.es}

\begin{abstract}
The cataclysmic variable SDSSJ~015543.40+002807.2 is confirmed to be a magnetic
system of the AM Herculis type.  With an orbital period of 87.13 minutes, it is
also the shortest-period eclipsing Polar known.  Monitoring with {\it
XMM-Newton\/} finds a high-state light curve dominated by a single X-ray
emitting accretion pole located slightly prograde of the secondary star.  The
hard X-ray spectrum is typical of radial shocks on magnetic white dwarfs
($kT\sim10$~keV), and there is evidence for a soft X-ray component consistent
with reprocessing from the stellar surface.  The optical circular polarization
is weak ($v\lesssim3\%$) when the accretion rate is high ($m_V\sim15.5$), due
to optically-thick cyclotron emission and the apparent competition between two
accreting poles.  However, in low states ($m_V\sim18$), the polarization
increases smoothly to the blue, reaching 20\% at 4200~\AA, and the flux
spectrum displays a rich set of thermally-broadened cyclotron harmonics that
indicate a polar field of 29~MG.  The phase interval preceding the 6.5~min
eclipse depicts the development of P-Cygni components followed by complete
absorption reversals in the emission lines.  This phenomenon is not unexpected
for a strongly accreting magnetic system viewed through the cool base of the
funnel, and high-quality spectroscopy through this interval will likely lead to
new insights into the dynamics of magnetic coupling and gas flow onto the white
dwarf.
\end{abstract}


\section{Introduction}

Despite a catalog of nearly 80 systems (e.g., Ritter \& Kolb 2003),
discoveries of new AM Herculis systems, or Polars, continue to provide insights
into the phenomenon of accretion onto compact magnetic objects through the
exploration of interesting geometries and new regions of parameter space.  For
example, we have only recently begun to appreciate the degree that a very
strong field ($B \gtrsim 100$~MG) on the white dwarf primary star reduces the
observed circular polarization -- the hallmark of strongly magnetic cataclysmic
variables (mCVs) -- by shifting the higher optical-depth, low-harmonic
cyclotron emission features into the optical portion of the spectrum (Schmidt
et al. 1996; 2001).  Unusually high specific mass-transfer rates ($\dot m
\gtrsim 10$~g~cm$^{-2}$~s$^{-1}$) can depress the accretion spot below the
surrounding photosphere, and thereby also yield low polarization (e.g.,
Stockman 1988).  By contrast, very low-$\dot m$ systems
($<10^{-2}$~g~cm$^{-2}$~s$^{-1}$) have begun to appear in deep spectroscopic
surveys (Reimers, Hagen, \& Hopp, 1999; Reimers \& Hagen 2000; Szkody et al.
2003), where the reduced Doppler broadening and high contrast to underlying
sources of emission give rise to striking cyclotron harmonic features that can
be mistaken for quasar emission lines. Due to the low optical depth, circular
polarization in these features can exceed 70\%.

Eclipsing geometries are particularly valuable for all binary types because
the transition timings and profiles allow the locations and sizes of emission
regions to be mapped within the systems.  The accretion binary
SDSS~J015543.40+002807.2 (hereafter SDSS~J0155+0028) was discovered by Szkody
et al. (2002) in the Early Release Data of the Sloan Digital Sky Survey
(SDSS).  The SDSS spectrum was saturated due to its brightness, but follow-up
time-resolved spectroscopy with the Apache Point Observatory in November 2000
showed strong \ion{He}{2} $\lambda$4686 emission and a large amplitude radial
velocity curve with an orbital period of 87 min. The most intriguing aspect of
the system was that the Balmer and \ion{He}{1} (but not the \ion{He}{2})
emission lines went into absorption for part of the orbit, while the continuum
did not change during those phases. In addition, the phasing of the absorption
did not correspond to the spectroscopic zero-crossing of the radial velocities,
such as would result from an eclipse of the white dwarf by the secondary star.
SDSS~J0155+0028 was also weakly detected in {\it ROSAT\/} survey observations.
Based on these characteristics, the system was proposed to be a likely high
inclination mCV.  If found to be strongly polarized, SDSS~J0155+0028 would be
the shortest-period eclipsing AM Her system known.

Optical photometry obtained when the system was bright ($m_V\approx15.2$) by
students at the Nordic-Baltic Research Course (Dubkova, Kudryavtseva \& Hirv
2003) in 2002 August supported this picture by revealing a deep (5.9 mag),
narrow ($\sim$400~s) eclipse superposed on a broad hump.  This is
characteristic of a magnetic accretion system dominated by a single pole in
which the accretion spot is self-eclipsed by the white dwarf for less than half
the orbital period.  Further photometry of SDSS~J0155+0028 by Woudt, Warner, \&
Pretorius (2004) on 2003 October 4$-$5 found the system in a low accretion
state ($m_V \sim 18$) but still strongly variable on a period of $87.13\pm0.02$
min. The brightness hump has a different shape and is narrower in width than in
the high state observations.

In this paper we report the results of {\it XMM-Newton\/} X-ray and
ground-based optical photometry, spectroscopy and spectropolarimetry that
confirm SDSS~J0155+0028 to be a magnetic CV. The system is characterized by
wide swings in mass-transfer rate and an unusual viewing geometry that offers
new avenues for studying the structure of accretion columns in Polars.

\section{Observations}

\subsection{{\it XMM-Newton} Observations}

X-ray observations of SDSS~J0155+0028 were obtained in a single time series
with the {\it XMM-Newton\/} Observatory on 2003 July 4 covering 14:27$-$16:34
UT.  Useful data were obtained by each of the EPIC X-ray CCD cameras (EPIC-pn,
and the two EPIC-MOS, each of which has about half the effective area of the
pn; Turner et al. 2001), using the ``Thin1'' filter and ``Prime Full Window''
mode.  These data are summarized in Table 1. However, the source was
sufficiently faint that neither the continuum nor any bright emission lines
were detected with the dispersing reflection grating spectrograph (RGS; den
Herder et al. 2001). Moreover, due to a technical failure, no usable
time-resolved data were available from the optical monitor (OM; Mason et al.
2001).  The data were analyzed following the ABC guide from the {\it
XMM-Newton\/} US GOF web
site\footnote{http://heasarc.gsfc.nasa.gov/docs/xmm/abc/abc.html} and analysis
threads from the main Vilspa
site\footnote{http://xmm.vilspa.esa.es/external/xmm\_sw\_cal/sas.shtml},
employing calibration files current to 2004 March 23 and the SAS v6.0. First,
the ODF files were reprocessed to produce new event files.  The event list was
then screened using the standard expressions and energies were restricted to
the range 0.1$-$10~keV.  For the pn, source data were extracted within a 360
pixel radius circular aperture (this only encircles $\sim70\%$ of the energy,
but for low count-rate sources reduces the background contribution) using a
background computed from adjacent rectangular regions at similar detector Y
locations to the target.  As advised, a conservative choice of event selection,
pattern = 0, was adopted.  For the two MOS detectors, event pattern $\leq$12
was applied and a 320 pixel aperture size was used (also enclosing $\sim$70\%
of the energy), but the background was based on an annulus surrounding the
source.  We note that the fluxes and count rates quoted in what follows have
been corrected for the encircled energy fraction.

Light curves were extracted for both source and background with the SAS task
{\tt evselect}, using the same extraction regions as the spectra, but with a
less conservative pattern $\leq4$ for the pn events.  The task {\tt lccor} was
then used to scale the data and achieve background subtraction, dead-time, and
vignetting corrections, producing net source light curves with 20~s and 100~s
binning.  Finally, the time stamps were corrected to the solar system
barycenter using FTOOLS tasks (giving HJD(TB)), and then 64.184~s was
subtracted to yield HJD(UT) for comparison with the ground-based observations.
The 100~s binned light curves for the three detectors are shown in Figure 1.
From the PIMMS conversion software available at the NASA HEASARC web
site\footnote{http://heasarc.gsfc.nasa.giv/Tools/w3pimms.html}, and an input
model consisting of an absorbed ($n_H=2.73\times10^{20}$~cm$^{-2}$), 7~keV
thermal bremsstrahlung component, the {\it XMM\/} peak count rates of
0.3~c~s$^{-1}$ in the MOS telescopes and 0.8~c~s$^{-1}$ in the PN (all
$\sim$0.1$-$10~keV) correspond to $\sim$0.09~cps with the {\em ROSAT} PSPC,
suggesting that the system was in an active accretion state, comparable to to
that during the {\it ROSAT\/} observation (0.04~cps).

\subsection{Ground-Based Observations}

Optical photometry was obtained at the United States Naval Observatory (USNO)
1~m telescope with a CCD and $V$ filter on the nights immediately preceding
(July 4 UT) and following (July 5 UT) the {\it XMM\/} observations, as well as
a long sequence on 2002 November 15. Differential photometry was used to
construct the light curves shown in Figure 2, which are folded and phased based
on the ephemeris described below.  While the results indicate that
SDSS~J0155+0028 may have been $\sim$1 mag fainter than when measured by
Dubkova et al. (2003), they confirm that the system was actively accreting
during the period of X-ray monitoring.

Additional CCD photometry during the high state was added to study the eclipse
in different filters and to establish the eclipse ephemeris.  Differential
photometry with respect to the same comparison star listed in Dubkova et al
(2003) was obtained with the Manastash Ridge Observatory 0.76~m telescope on
2002 September 14 using a $V$ filter and 15~s exposure times. Differential
photometry with a combination of comparison stars using the 2.56~m Nordic
Optical Telescope (NOT) on 2002 October 16 provided $B$ light curves with 8.2~s
integrations and an $R$ light curve with 7.8~s integrations. On that same
night, 30~s integrations without a filter were obtained at the 0.6~m telescope
at the Observatorio del Roque de los Muchachos. Further $B$ observations using
27.9~s integrations were also made at the 0.8~m telescope at Observatoria del
Teide on 2002 November 14. Figure 3 shows representative light curves in $B$,
$V$ and $R$ filters from these observations and the eclipse timings are given
in Table 2. The light curves are fairly similar in the different filters. Only
the $R$ filter has the time resolution and signal-to-noise ratio to provide
some information on the eclipse shape, indicating that it does not appear to
have a flat bottom and the egress is steeper than the ingress. Comparing this
profile with the $V$ data of Dubkova et al (2003) (both obtained with the NOT),
it appears that the eclipse in $V$ is about 2 magnitudes deeper than in $R$.
The other point of note is the presence of a brief but noticeable drop (almost
a magnitude) about 0.1 phase prior to the eclipse, which is a prominent
feature in $B$ and $R$ (but less so in $V$) in all data from 2002 August
through November. If this feature arises through absorption by the stream (see
$\S$5), it is not a permanent feature as it is not apparent in the high state
data of 2003 July (Figure 2).

Optical spectrophotometry and circular spectropolarimetry of SDSS~J0155+0028
were obtained on 2002 December 5 UT with the 2.3~m Bok telescope of Steward
Observatory and again on 2004 February 16 UT with the 6.5~m MMT atop Mt.
Hopkins.  Both runs made use of the CCD SPOL spectropolarimeter (Schmidt,
Stockman, \& Smith 1992), configured for broad spectral coverage
($\lambda\lambda$4200$-$8200) at a rather low resolution of $\sim$15~\AA.  In
2003 the instrument was upgraded with a cosmetically perfect,
1200$\times$800$\times$15~$\mu$m pixel CCD from Semiconductor Technology
Associates, Inc. that was thinned, antireflection-coated, and permanently
sensitized through backside charging at the Steward Observatory Imaging
Technology Laboratory.  The time series spectropolarimetry used an exposure of
240~s per observation for an effective time resolution of
$\Delta\varphi\sim0.065$. Total flux spectra are recorded at twice this
frequency. The 2002 Dec. observations found SDSS~J0155+0028 in a high accretion
state, with $m_B=15.1$ at maximum brightness and $\sim$20 at mid-eclipse.  This
data series is also the more extensive, covering 1.5 orbital periods (and 2
eclipses). The 2004 Feb. observations were obtained as the object set in
evening twilight, so were limited to only slightly more than a quarter of the
orbital period. From the comparison of spectrum-added circular polarization and
$B$-band brightness shown in Figure 4, it is clear that SDSS~J0155+0028 was in
a much lower accretion state at this epoch, with $m_B>18$ and fading through
the sequence. Again, phasing was accomplished using the ephemeris derived
below.  Note the general similarities between the two panels of Figure 4,
notably the rapid drop in brightness between $\varphi=0$ and 0.3 that has been
seen at all epochs to date (e.g., Figures 2 and 3; Dubkova et al. 2003; Woudt
et al. 2004), and the same sign of circular polarization in this phase interval
(see also below).

\section{High State X-ray Light Curve and Spectrum}

Both the X-ray and contemporaneous optical photometric light curves show a
single bright phase, punctuated near the middle by a virtually total eclipse of
duration $\sim$390~s ($\Delta\varphi\sim0.075$).  This is best seen in the
light curve obtained by summing the two MOS detectors and folding on the
orbital period, as shown in Figure 5.  Referenced to the middle of eclipse, the
bright phase appears as a smooth ``hump'' lasting from $\varphi=0.6$ to 0.3,
i.e., commencing $\sim$0.1 in phase prior to the optical, but terminating at
the same phase as the optical after the eclipse. The rise toward peak
brightness in X-rays is also steeper than in the optical; indeed X-ray maximum
occurs slightly ahead of eclipse, whereas the optical light doesn't peak until
$\Delta\varphi\sim0.1$ after eclipse.  Within the uncertainties, the profiles
of the bright phases and eclipses are consistent over the 2.4 cycles covered.
Light curves extracted over different energy intervals (e.g., $0.1-0.6$~keV and
$0.6-10$~keV) are strongly signal-to-noise ratio limited and do not provide
useful information on hardness ratio changes over interesting phase intervals.

In the standard picture for Polars in their active accretion state ($\dot m
\gtrsim 1$~g~cm$^{-2}$~s$^{-1}$) the X-ray emission is characterized by two
components: (i) a hard ($kT\sim10-30$~keV) thermal plasma component (e.g.
bremsstrahlung) from the post-shock accretion column, and (ii) a soft,
blackbody component from the underlying irradiated white dwarf photosphere.
Theoretically, the reprocessed fraction of the total accretion energy is
expected to be $L_{bb}/L_{acc}\sim0.56$ (King \& Watson 1987) . However, X-ray
observations of Polars have often indicated a so-called ``soft X-ray excess''
(Ramsay et al. 1994), with the soft/hard ratio in excess of one, hence
additional heating mechanisms have been proposed to enhance the relative
contribution from the blackbody.  The most widely accepted is the
``bombardment'' scenario, wherein the accretion funnel contains blobs of plasma
of varying lengths/densities, the densest being able to penetrate beyond an
optical depth of unity into the white dwarf before being shocked, effectively
heating the atmosphere from below (Kuijpers \& Pringle 1982).

We chose a single temperature bremsstrahlung model for the harder component,
as the aim of this relatively short observation was to place constraints on the
temperatures and hard/soft flux contributions.  A joint fit of a blackbody plus
thermal bremsstrahlung was performed on the three PN and MOS datasets within
XSPEC\footnote{http://heasarc.gsfc.nasa.gov/docs/xanadu/xspec/index.html}.  The
data were binned to give $>$30 counts per bin and $\chi^2$ statistics were used
to find the best overall fit to the background subtracted spectra. There remain
uncertainties in the cross-calibration of the effective areas of the PN and MOS
detectors, hence the model normalizations were allowed to differ. Fits
attempted with both column density $n_H$ and $kT_{bb}$ as free variables were
not well constrained, therefore fits were examined for typical values of
$kT_{bb}=20$, 30 and 40~eV, and $n_H$ ranging from
$0.5-2.7\times10^{20}$~cm$^{-2}$, the upper limit being the value given by the
HEASARC column density
tool\footnote{http://heasarc.gsfc.nasa.gov/cgi-bin/Tools/w3nh/w3nh.pl}, which
estimates the column through the entire Galaxy for this sky position.

The reliability of the detector calibrations at the lowest energies is still
unclear. We therefore used two sets of low energy cut-offs when fitting the
spectra: i) energies $>$0.15~keV for PN and $>$0.2~keV for MOS, potentially
useful for constraining any soft blackbody component, and ii) more conservative
limits of $E>0.3$~keV (PN) and $>$0.5~keV (MOS). Including the lowest energy
bins we find equally good fits ($\chi^2_{\nu}=1.1$, 101 degrees of freedom)
with $n_H=0.5-1.0\times10^{20}$~cm$^{-2}$, $kT_{br}=8.0^{+1.6}_{-1.0}$~keV and
any of the blackbody temperatures.  However, only in the case of the coolest
blackbody and $n_H=1.0\times10^{20}$~cm$^{-2}$ is the soft component required
for the fit (an $F$-test yields a 98\% confidence level, compared to $\ll1$\%
for the other temperatures).  This 20~eV blackbody contributes
$1.1\times10^{-12}$~erg~cm$^{-2}$~s$^{-1}$ to the unabsorbed $0.01-10$~keV
flux, while in all cases the bremsstrahlung component provides
$2\times10^{-12}$~erg~cm$^{-2}$~s$^{-1}$. On the other hand, if $n_H$ is as
high as $2.7\times10^{20}$~cm$^{-2}$, for all blackbody temperatures the
$F$-test gives confidence levels $>$95\% for inclusion of the soft component,
and the best fit is obtained for $kT_{bb}=40$ eV and $kT_{br}=6.7\pm0.8$~keV.
This fit, which also yields $\chi^2_\nu=1.1$ for 101 d.o.f., is shown in Figure
6. Again the bremsstrahlung flux is $2\times10^{-12}$~erg~cm$^{-2}$~s$^{-1}$,
with the blackbody adding a further
$0.6-26\times10^{-12}$~erg~cm$^{-2}$~s$^{-1}$ for the $40-20$~eV temperatures,
respectively.  The more conservative lower limits do not lead to a constraint
on the blackbody component at all.  This is almost certainly a consequence of
losing the information below 0.3~keV. For a thermal bremsstrahlung component
alone we find temperatures of $7.8\pm1.0$ and $6.5\pm0.8$~keV for column
densities of 0.7 and $2.73\times$10$^{20}$~cm$^{-2}$, respectively, with the
same flux as above.

Accounting for geometrical effects, the X-ray scattering albedo of the white
dwarf photosphere, $a_x=0.3$ (Williams, King, \& Brooker 1987), and neglecting
the cyclotron term (usually an order of magnitude smaller than the
bremsstrahlung component) gives luminosities of $L_{bb}\approx\pi F_{bb}(1-a_x)
d^2$ and $L_{acc}\approx L_{br}=2\pi F_{br}(1+a_x) d^2$, where $d$ is the
distance.  This leads to a soft/hard luminosity ratio of $0.1-3.5$ for the
higher $n_H$ value ($kT_{bb}=40-20$~eV) or $0-0.15$ for the lower values of
$n_H$. It seems more likely that column density lies in the range
$1.0-2.7\times10^{20}$~cm$^{-2}$, as the absence of a soft component is
difficult to explain given an active accretion state at the time.

The low ratio of soft/hard X-ray luminosity is as expected for the standard
radial accretion model for Polars, although we cannot rule out a soft X-ray
excess if the blackbody is sufficiently cool.  Indeed, the most recent synopsis
of {\em XMM} results on 21 magnetic systems by Ramsay and Cropper (2004) has
cast doubt on the existence of the soft X-ray excess among Polars in general.
Calibration issues as discussed above complicate a good determination for the
soft components, and corrections must be made for reflection of the hard
X-rays, optical thickness, and the contributions from the cyclotron emission.
With these considerations in mind, we conclude that SDSS~J0155+0028 is not
unusual among Polars from an X-ray standpoint.

\section{A Short-Period Magnetic CV with Wide Variations in Accretion Rate}

\subsection{Orbital Ephemeris}

Optical photometry of SDSS~J0155+0028 by Woudt et al. (2004) led to a
determination for the orbital period of $87.13\pm0.02$~min, or
$0.060507(14)$~d.  Here, the numbers in parentheses indicate the uncertainty in
the final digits.  We have combined the results of Woudt et al. with the
eclipse timing of Dubkova et al. (2003) and the new X-ray and optical timings
reported here.  Table~2 summarizes all 19 eclipses spanning more than a year,
with times of mid-eclipse shown as heliocentric Julian dates (HJD).  The
measurements are distributed such that a common orbital ephemeris can be
derived without cycle-count ambiguity. The result, weighted by the estimated
uncertainties of individual timings, and referenced to the second eclipse
observed on 2002 December 5, is
\begin{equation}
{\rm HJD} = 2452613.78907(11) + 0.06051621(4)\cdot E
\end{equation}
For convenience, the cycle count appropriate to each eclipse is also listed in
Table~2, and an ``O-C'' (observed minus computed) diagram is shown for the
eclipse timings in Figure 7.  The period is 2.7~min shorter than that of the
previously shortest-period Polar, DP Leo.  The accuracy of the ephemeris is
sufficient to phase future observations of SDSS~J0155+0028 to better than
$\Delta\varphi=0.05$ for a decade.

\subsection{Changes with Accretion State}

The sharp contrast in spectral properties between high and low accretion
states of SDSS~J0155+0028 is illustrated by excerpts from the time series
spectropolarimetry shown as the bold curves in the top and bottom panels of
Figure 8, respectively. Both panels represent the system soon after maximum
light ($\varphi=0.1$).

The high state in 2002 December is distinguished by a very blue continuum and
emission-line spectrum showing H, \ion{He}{1}, and \ion{He}{2} lines, with the
Balmer decrement H$\alpha$\,:\,H$\beta$\,:\,H$\gamma$ inverted and
$F_{\lambda4686} \approx F_{H\beta}$.  The line ratios are characteristic of
an accretion funnel in a highly active system, and any line emission component
from the inner hemisphere of the heated secondary cannot be resolved in these
low-resolution data.  Cyclotron emission harmonics in the high state are
indistinct, but peaks can be recognized near 5600\AA, 6550\AA, and 7600\AA.
The circular polarization is weak: +3\% overall, increasing mildly to the
blue. By contrast, the low-state flux spectrum from 2004 February displayed in
the bottom panel is roughly flat in $F_\lambda$ and shows emission lines that
have faded in approximate proportion to the continuum. \ion{He}{1} emission is
nearly absent and $F_{\lambda4686} < F_{H\beta}$. A rich spectrum of
well-defined cyclotron features is evident in the low state, with peak
wavelengths at $\sim$5180\AA, 5590\AA, 6210\AA, 6860\AA, and 7780\AA.  The net
circular polarization increases monotonically from only a few per cent at the
long-wavelength extreme to nearly 25\% in the blue.

It is straightforward to interpret the highly structured, strongly polarized
continuum in the low state as cyclotron emission from primarily a single
accretion spot.  Fitting the peak wavelengths in the above list to the formula
for cyclotron emission in a single-temperature plasma (Cropper et al. 1988;
Wickramasinghe 1988) results in the assignment of harmonic numbers $m=10-6$,
respectively, a magnetic field strength in the spot of 29~MG, and a temperature
factor $kT\sin^2\theta=15$~keV. The increase in polarization with frequency is
a natural result of the rapid decline in cyclotron absorption coefficient with
increasing harmonic number (Chanmugam et al. 1989).  Note that even
though we characterize 2004 Feb. as a low state, the optical depth at $m=6$
(7800\AA) is still high enough to quench the circular polarization, the
harmonics overlap considerably, and the accretion column is optically thick in
the Balmer series.  Indeed, the flux spectrum in the bottom panel of Figure 8
is reminiscent of the low-state spectrum of VV Puppis ($B\sim30$~MG;
Wickramasinghe \& Visvanathan 1979), where the accretion rate has been
estimated to be at least an order of magnitude higher than for chronically
anemic Polars like SDSS~J1553+5516 ($<$$10^{-13}~M_\sun$~yr$^{-1}$; Szkody et
al. 2003).

With an order of magnitude increase in brightness and circular polarization
that is of the same sign but much reduced in amount, it would appear that the
2002 Dec. high state of SDSS~J0155+0028 could be explained simply by
optically-thick emission from an increased mass transfer rate onto the same
magnetic pole. However, the diffuse cyclotron emission bumps evident at this
epoch are more widely spaced, indicative of the addition of a second active
accretion spot with a different field strength. Unfortunately, only three
diffuse harmonics do not lead to a definitive solution, but the best fit
occurs for an assignment of $\lambda\lambda$5600, 6550, 7600, to $m=6,5,4$
respectively, $B\sim48$~MG, and $T\sin^2\theta\sim30$~keV.  This suggests that
the reduction in polarization in the high state may be due, in part, to the
competition of two emission regions with opposite polarity.  Simultaneous
visibility of two accretion spots with unequal field strengths could arise if
they represent complementary footpoints of closed field lines in an offset
dipolar structure (e.g., Wickramasinghe, Ferrario, \& Bailey 1989).

\subsection{System Parameters}

The eclipse duration in SDSS~J0155+0028 has not yet been measured with high
precision (our best time resolution is 10~s in the NOT observations).
Nevertheless, available estimates permit useful checks on key parameters of
the binary.  For example, the generally high level of activity suggests that
the secondary star fills its Roche lobe, which for a binary with $P=87$~min
implies a radius $R_2=0.14$~$R_\sun$ (e.g., Pringle \& Wade 1985). The maximum
predicted eclipse length of an emission region on the white dwarf (as seen by
an observer in the plane of the orbit) is a function of white dwarf mass,
ranging from 460 to 380~s for $M_1=$ 0.4 to 0.8~$M_\sun$, respectively.
Estimates for SDSS~J0155+0028 range from $\sim$320~s (Woudt et al. 2004) to
$\sim$390~s for our {\it XMM\/} data. Because a non-central eclipse reduces
the duration of totality, these values favor the lower end of the mass range.
If the secondary has the structure of a main-sequence star (questionable given
the advanced age of this binary, but see Beuermann et al. 1998), it has a mass
$M_2=0.11$~$M_\sun$ (Beuermann \& Weichhold 1999), an absolute visual
magnitude $M_V\sim15.1$ (Baraffe et al. 1998), and a spectral type of M5.5
(Henry, Kirkpatrick, \& Simons 1994). A distance estimate is then possible by
scaling the weak rise in spectral flux beyond $\lambda=7000$~\AA\ that is
evident when the accretion spot is hidden from view to library flux spectra
for M dwarfs.  The best data for this purpose come from the low-state,
faint-phase ($\varphi=0.38$) MMT spectrum shown in the bottom panel of Figure
8, and we use for comparison the M dwarf standards that have been placed on an
absolute magnitude scale in Sloan bands by (Hawley et al. 2002). We find
$D\sim290$~pc, with an uncertainty of at least 25\%. However, even this crude
estimate yields reasonable numbers for a Polar in an active accretion state
(cf. Warner 1995): a $0.01-10$~keV luminosity of
$L_X\sim2\times10^{31}$~ergs~s$^{-1}$ and an accretion rate $\dot
M>4\times10^{-12}$~$M_\sun$~yr$^{-1}$, the inequality resulting from the
likelihood that cyclotron emission and the largely unmeasured soft X-ray
component carry a significant fraction of the accretion luminosity. Compared
with the several dozen Polars that have been studied over the past three
decades, it appears that SDSS~J0155+0028 is largely unremarkable except for an
unusual viewing perspective, as discussed below.

\section{Absorption in the Accretion Funnel}

Eclipse profiles are rich sources of information on the locations, sizes, and
properties of emission regions in magnetic CVs.  The overall duration of
eclipse reflects the chord executed by the white dwarf and accretion spot(s)
across the secondary star, and for a typical geometry lasts
$\Delta\varphi\sim0.08$. Ingress and egress of the white dwarf disk each
require at least 30~s, depending on primary star mass, orbital period, and
inclination, and appear most prominently in the near-UV (Schmidt \& Stockman
2001). In the optical, the white dwarf is usually overwhelmed by the
combination of the tiny accretion spot(s) (e.g., Bailey \& Cropper 1991;
Schwope et al. 2001) and continuum emission from the funnel, each of which
provides a characteristic eclipse signature (e.g., Stockman \& Schmidt 1996).
The accretion stream can also be detected through absorption dips preceding
primary eclipse in the emission lines and EUV (Schwope et al. 2001). For
SDSS~J0155+0028, the photometric database stands to be improved in both
wavelength coverage and time resolution, but the eclipse depth ($m_V\sim20$ at
minimum) implies that a large telescope will be required.

Despite the limitations of available data, new information on the properties
of the accretion stream in the high state -- including the origin of the line
absorption phase described by Szkody et al. (2002) -- is available from our
complete orbit of spectrophotometry obtained at the 2.3~m telescope in 2002
December.  In Figure 9 we display the radial velocity and flux in the
H$\alpha$ emission line as a function of orbital phase for all phases where
the line is uncontaminated by absorption. The results confirm that the
accretion stream and funnel are the dominant sources of line emission in this
state.  Not only are the $\gamma$ velocity and line widths very large
(+200~km~s$^{-1}$; 1000$-$1500~km~s$^{-1}$, respectively), but the velocity
curve is somewhat triangular, shows maximum redshift near the time of eclipse,
and the line flux possesses two minima: one at eclipse ($\varphi=0$) and a
second 0.5 orbit later. All of these are characteristics of optically-thick
emission in a flow that is directed largely along the line of sight near the
stellar conjunctions.

The spectral sequence for the interval around eclipse is presented as Figure
10.  Phase proceeds downward as marked, in intervals of 2.6~min
($\sim$11$^\circ$ of orbital motion).  The initial two spectra, taken at
$\varphi=0.82$ and 0.85, show strong Balmer, \ion{He}{2}, and \ion{He}{1}
emission lines typical of active AM Her systems.  The lines at this phase are
approaching maximum redshift and are marginally resolved with a full width at
half-maximum (FWHM) of $\sim$1000~km~s$^{-1}$.  At $\varphi=0.88$, 40$^\circ$
prior to the primary mid-eclipse, the lines suddenly fade and P-Cygni type
absorption components develop $\sim$450~km~s$^{-1}$ to the blue of the emission
peaks in all lines {\it except\/} \ion{He}{2} $\lambda$4686. By $\varphi=0.91$
(30$^\circ$ prior to mid-eclipse), the absorption features have deepened and
shifted redward to completely consume the emission lines.  This is the phase
of the photometric dip discussed in $\S$2.  The prominence of the dip in $B$
and $R$ (Figure 3) as opposed to $V$ reflects the presence of H$\beta$ and
H$\alpha$ in the two former filter bands.  Resuming the time series, the
absorption lines narrow to $\sim$800~km~s$^{-1}$ FWHM and continue to shift
some 400~km~s$^{-1}$ to the red before the P-Cygni phase returns at
$\varphi=0.97$. Following this, the system quickly drops into the deep primary
eclipse. This transformation is better displayed in an enlarged version of the
phase interval shown as Figure 11.

At mid-eclipse, the emission lines are reduced in brightness by about an order
of magnitude.  The weak continuum has $m_V\sim20$ and exhibits a somewhat concave
appearance.  Because the white dwarf is eclipsed, the continuum at this time is
probably bound-free and free-free emission (e.g., Kim \& Beuermann 1996) plus
possibly scattered cyclotron emission from the portion of the accretion stream
and funnel that is still visible. Eclipse egress occurs over the course of
$\Delta\varphi\sim0.06$ ($\Delta\theta\sim20^\circ$), with full exposure of
the line-emitting funnel lagging the continuum.

We recognize the development of P-Cygni components followed by full absorption
reversals of the lines for $0.88 < \varphi < 0.97$ as a self-eclipse of the
accretion funnel, a rare opportunity afforded by our fortuitous viewpoint of
SDSS~J0155+0028.  Only EF Eri (Verbunt et al. 1980) and MN Hya (Ramsay \&
Wheatley 1998) have shown this effect, and then in only a few lines.  The
geometry for SDSS~J0155+0028 is depicted in the sketch shown as Figure 12.
Here, the Earth lies in the plane of the paper, with the observer orbiting the
binary in a clockwise direction.  The stellar components are shown to scale for
an assumed 0.6~$M_\sun$ white dwarf and Roche-lobe filling secondary. Also
indicated is the ballistic stream trajectory and viewing phases corresponding
to the spectra displayed in Figure 10. The shaded area in the figure is a
schematic representation of the accretion funnel that is consistent with the
observations.  In order that a self-eclipse occurs, the funnel must terminate
in one or more accretion spots not far from the white dwarf equator.  The
azimuth range of the funnel has been estimated from the duration of the
absorption-line phase, but we also note that with this geometry, full exposure
of the stream following primary eclipse of the white dwarf would not occur
until nearly $\varphi=0.1$, as observed.  The inference from this sketch is
that the dominant magnetic pole lies at a longitude of $\psi\approx+30^\circ$,
i.e., that it leads the secondary in its orbit.  This is supported by the
folded X-ray light curve in Figure 5, where it can be seen that the onset and
conclusion of the bright phase are displaced around the deep eclipse by
$\Delta\varphi \approx -0.1$.

It was realized early in the studies of AM Herculis (Stockman et al. 1977) that
the high-state emission lines of Polars are formed in dense ($N_e \gtrsim
10^{13}$~cm$^{-3}$), collimated inflows that are optically thick in the
permitted lines.  Heating to temperatures of $\sim$10,000$-$40,000~K is
provided largely by hard and soft X-ray irradiation from the vicinity of the
shock, with the temperature generally falling with radial distance (e.g.,
Ferrario \& Wickramasinghe 1993).  It is therefore not surprising that the
lines could be seen in absorption if the funnel were viewed through its cool,
upper end.  However, models of the emission lines in mCVs constructed by
Ferrario \& Wehrse (1999) found that, while radiative heating alone can account
for the Balmer lines, it is insufficient to explain the strengths of
\ion{He}{1} and \ion{He}{2} lines.  The large volume present in the coupling
region -- where the gas attaches to the white dwarf's magnetic field lines --
offered an attractive alternative site for the origin of the helium features,
but an additional source of heating was required.  Magnetic turbulence and
reconnection, or weak shocks were suggested as candidate mechanisms.  These
ideas make certain predictions that can be tested with our observations of
SDSS~J0155+0028.

Referring again to Figure 11, we note first that the H and \ion{He}{1} lines
behave similarly through the spectral sequence, indicating no qualitative
difference in their origins.  Given the contrasting behavior of \ion{He}{2}
$\lambda$4686 (see also below), this is already at odds with the predictions of
Ferrario \& Wehrse (1999).  P-Cygni absorption components initially appear in
the blue wings of the Balmer and \ion{He}{1} lines, confirming that the
foreground gas is both cooler and receding with a smaller velocity than what
causes the bulk of the line emission.  This situation is reproduced at the
termination of self-eclipse near $\varphi = 0.97$. As the funnel self-eclipse
progresses, absorption components deepen and move toward redder wavelengths.
The fact that the phase of maximum line depth ($\varphi \approx 0.9$) coincides
with our most face-on view of the accretion spot indicates that the funnel is
primarily radial, as indicated in Figure 12. It is interesting, however, that
the absorption components completely consume the emission lines.  Indeed, the
line width (FWHM $\approx$ 1400~km~s$^{-1}$ at maximum depth) appears to exceed
that of the preceding emission lines ($\sim$1000~km~s$^{-1}$).  This might be
understood if the funnel temperature increases not only as one moves toward the
white dwarf, but also from the funnel core to surface.  Such a profile would,
of course, result from radiative heating from below.  A sightline through the
shielded funnel base, such as we are afforded at $\varphi = 0.9$, would then
view comparatively cool gas projected against the hot funnel skin over the full
range of radial velocities, and very broad absorption lines would result. The
sketch for SDSS~J0155+0028 shown in Figure 12 corresponds closely with a
similar diagram showing an accretion curtain and vertical stream in HU~Aqr
(Schwope, Mantel, \& Horne 1997).

The progressive shift of the absorption features to higher velocity through
the funnel eclipse phase is more puzzling, given the simple geometry sketched
in Figure 12.  Certainly gas that attaches to field lines early in its journey
(and so appears in absorption late in the sequence) does so onto paths that
are more closely aligned with their original ballistic stream trajectories;
this gas might therefore preserve a higher fraction of its infall velocity.
Alternatively, the gas density might vary with azimuth in such a way that we
see more deeply into the funnel at later eclipse phases, thus absorption
appears at higher redshift at these times.  Additional spectroscopy with both
higher time and velocity resolution would help to clarify among the
possibilities.

The \ion{He}{2} $\lambda$4686 emission line is nearly stationary in wavelength
and largely unchanged in brightness through the funnel self-eclipse, the
largest effect being the formation of a brief and weak P-Cygni component at
the phase of maximum absorption-line depth.  These properties are consistent
either with the funnel being essentially optically thin in $\lambda$4686,
and/or with the emission line being produced primarily in a coupling region
high above the white dwarf, as Ferrario \& Wehrse (1999) suggest.  We note that
the high-state data shown in Figures 10 and 11 were obtained with a 2.3~m
telescope, so better quality and resolution are not only desirable but
practical.

\section{Summary and Conclusions}

With a period of 87.13~min, SDSS~J0155+0028 is the shortest-period eclipsing
Polar known.  Observations to date indicate that the system is highly active
($m_V = 15-16$) a large fraction of the time, and reveal an X-ray light curve
and spectrum, as well as a low-state optical cyclotron emission spectrum, that
are characteristic of active Polars ($kT \sim 10$~keV; $B\sim30$~MG). The
$>$5~mag deep, $\sim$6.5~min duration eclipse has yielded an accurate
ephemeris, but photometry with higher time resolution and spectrophotometry
with better spectral resolution will provide a more detailed understanding of
the stellar components, the extent of the funnel, and the accretion spot
location(s) on the white dwarf surface.  Particularly interesting is an
absorption-line phase that immediately precedes the main eclipse.  This is
interpreted as the result of a favorable viewing perspective of the
optically-thick accretion flow through the cool base of the funnel, which must
lie virtually in the plane of the orbit. Existing observations have yielded a
simple description of the gas flow, but future high-quality spectrophotometry
through the absorption-line phase promises to be a powerful technique for
mapping the trajectory and properties of the accretion flow in this magnetic
CV.

\acknowledgements{The authors thank P. Woudt for sharing his photometry in
advance of publication.  GDS is grateful to the Director of Steward Observatory
for funding the foundry run of STA devices, from which an improved CCD for the
spectropolarimeter was obtained, and to M. Lesser for optimizing the chip for
astronomy.  The anonymous referee offered several useful suggestions and
corrections to an earlier version of the manuscript.  Support was provided by
NASA through {\it XMM\/} Grant NAG5-12938 and the NSF through AST 03-06080.
JES acknowledges support by the ``Access to the ENO'' project 02-038 from the
European Union. This work is based, in part, on observations obtained with {\it
XMM-Newton\/}, an ESA science mission with instruments and contributions
directly funded by ESA Member States and the USA (NASA).  Part of the optical
data were taken using ALFOSC, which is owned by the Instituto de Astrofisica de
Andalucia (IAA) and operated at the Nordic Optical Telescope under agreement
between IAA and the NBIfAG of the Astronomical Observatory of Copenhagen.}


\clearpage


\begin{deluxetable}{lcccl}
\tablewidth{4.25truein}
\setlength{\tabcolsep}{0.06in}
\tablecaption{{\it XMM-Newton\/} Observations of SDSS J0155+0028}
\tablehead{
\colhead{Instrument}
& \colhead{Date}
& \colhead{UT}
& \colhead{Exposure}
& \colhead{Counts s$^{-1}$} \\
& \colhead{(yyyy-mm-dd)}
& \colhead{}
& \colhead{(s)}
}
\startdata
PN  & 2003-07-04 & 14:27$-$16:34 & \ \,4920 & 0.0$-$1.00 \\
MOS1 & 2003-07-04 & 13:13$-$16:37 & 12254 & 0.0$-$0.28 \\
MOS2 & 2003-07-04 & 13:13$-$16:37 & 12264 & 0.0$-$0.34 \\
\enddata

\end{deluxetable}

\begin{deluxetable}{crl}
\tablewidth{3.truein}
\setlength{\tabcolsep}{0.06in}
\tablecaption{Eclipse Times for SDSS~J0155+0028}
\tablehead{
\colhead{Eclipse Center}
& \colhead{Cycle}
& \colhead{Source}
\\
\colhead{(HJD $-$ 2450000)}}
\startdata
2506.6751 & $-$1770 & ~~1, NOT 2.56~m \\
2531.9713 & $-$1352 & ~~2, MRO 0.76~m \\
2563.560\, & $-$830  & ~~2, NOT 2.56~m \\
2563.621\, & $-$829  & ~~2, NOT 2.56~m \\
2563.686\, & $-$828  & ~~2, IAC 0.6~m \\
2592.544\, & $-$351  & ~~2, IAC 0.8~m \\
2592.605\, & $-$350  & ~~2, IAC 0.8~m \\
2593.7586 & $-$331  & ~~2, USNO 1.0~m \\ 
2593.8194 & $-$330  & ~~2, USNO 1.0~m \\ 
2593.8788 & $-$329  & ~~2, USNO 1.0~m \\ 
2613.7281 & $-$1    & ~~2, SO 2.3~m \\
2613.7892 & 0       & ~~2, SO 2.3~m \\
2824.932\,  & 3489  & ~~2, USNO 1.0~m \\
2825.0513 & 3491    & ~~2, {\em XMM} \\ 
2825.1115 & 3492    & ~~2, {\em XMM} \\ 
2825.1717 & 3493    & ~~2, {\em XMM} \\ 
2825.960~\, & 3506  & ~~2, USNO 1.0~m \\
2917.3990 & 5017    & ~~3, SAAO 1.9~m \\ 
2918.4278 & 5034    & ~~3, SAAO 1.9~m \\ 
\enddata
\tablerefs{$^1$Dubkova et al. (2003); $^2$This paper; $^3$Woudt et al. (2004).}

\end{deluxetable}

\clearpage

\begin{figure}
\includegraphics{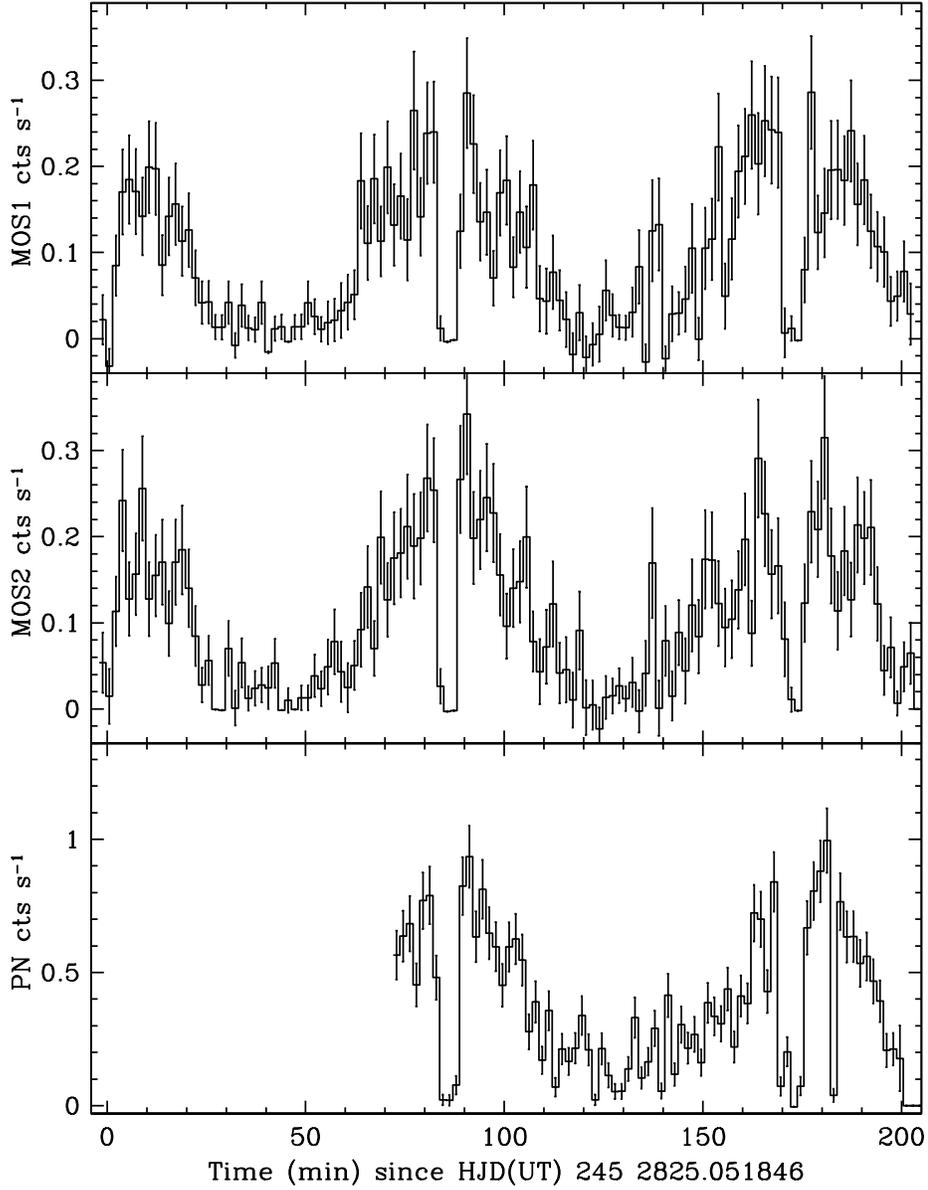}
\vspace{4.75truein}

\figcaption{{\it XMM-Newton}/EPIC-pn and MOS time series with 100~s binning.
Note the single-humped light curve characteristic of a one-pole accretor,
punctuated by a total eclipse of duration $\sim$390~s.  The latter half of
the PN light curve was obtained as a large number of short exposures, so
the exposure time per bin is nonuniform and occasionally very short.}

\end{figure}

\clearpage

\begin{figure}
\includegraphics{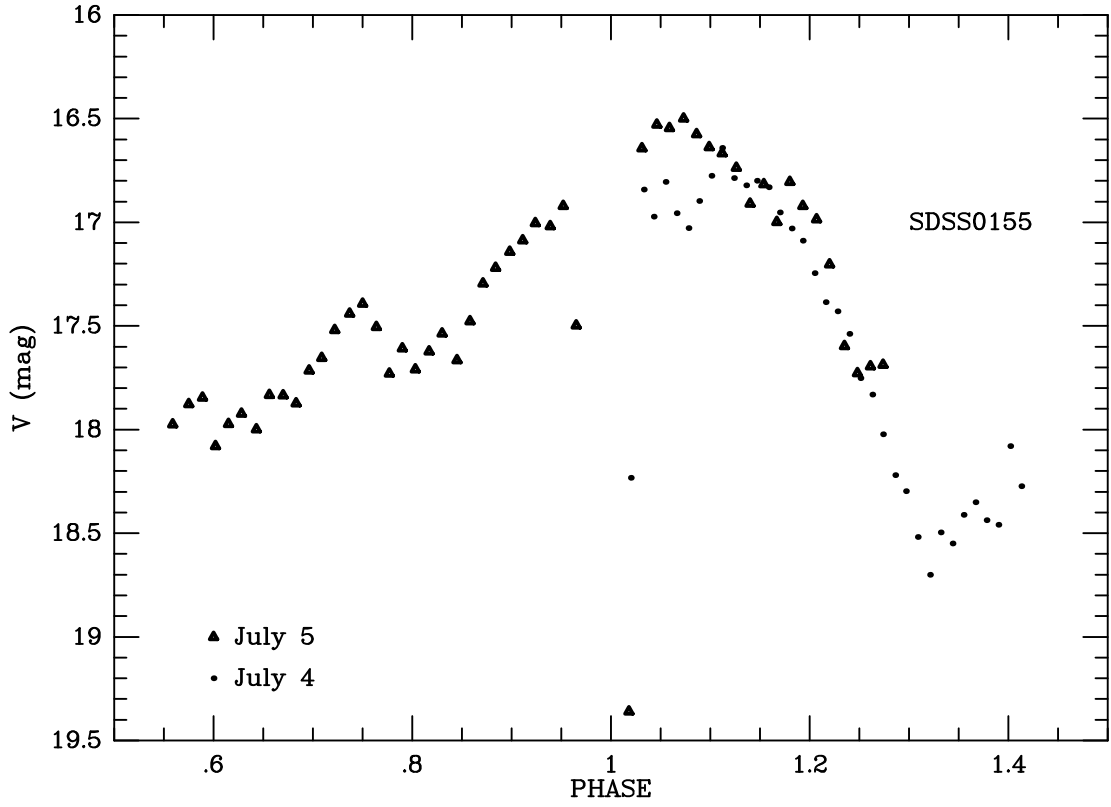}
\vspace{2.truein}

\figcaption{Optical photometry obtained on the nights immediately preceding
(July 4 UT) and following (July 5 UT) the {\it XMM\/} observations, confirming
that SDSS~J0155+0028 was in an active accretion state for the X-ray
measurements.}

\end{figure}

\clearpage

\begin{figure}
\includegraphics{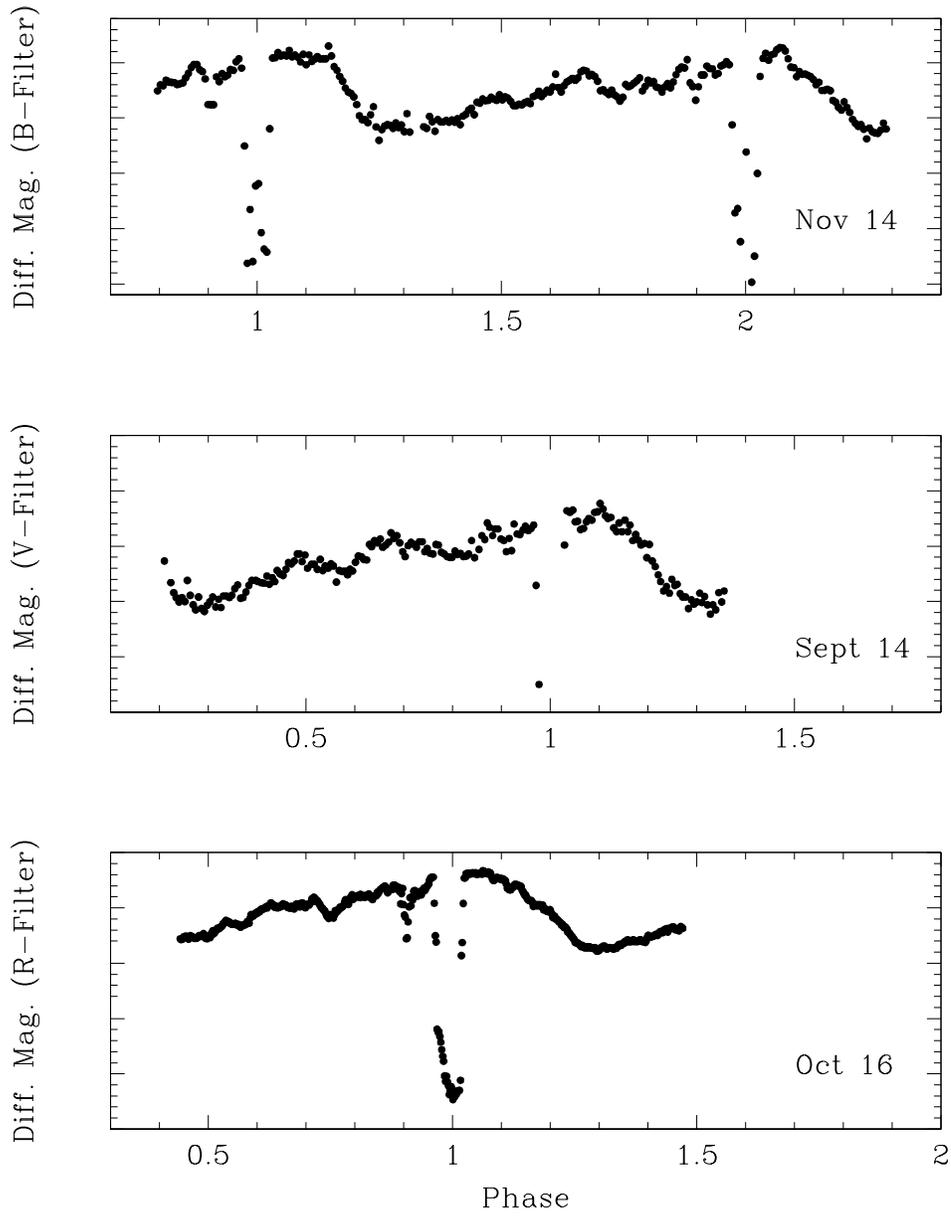}
\vspace{5.5truein}

\figcaption{Differential photometry of SDSS~J0155+0028 obtained during the high
accretion state with various filters and telescopes in the fall of 2002.  Note
the brief dip near $\varphi=0.9$ in each panel, suggestive of a stream
eclipse.  Each large interval on the ordinate represents a difference of one
magnitude.}

\end{figure}

\clearpage

\begin{figure}
\includegraphics{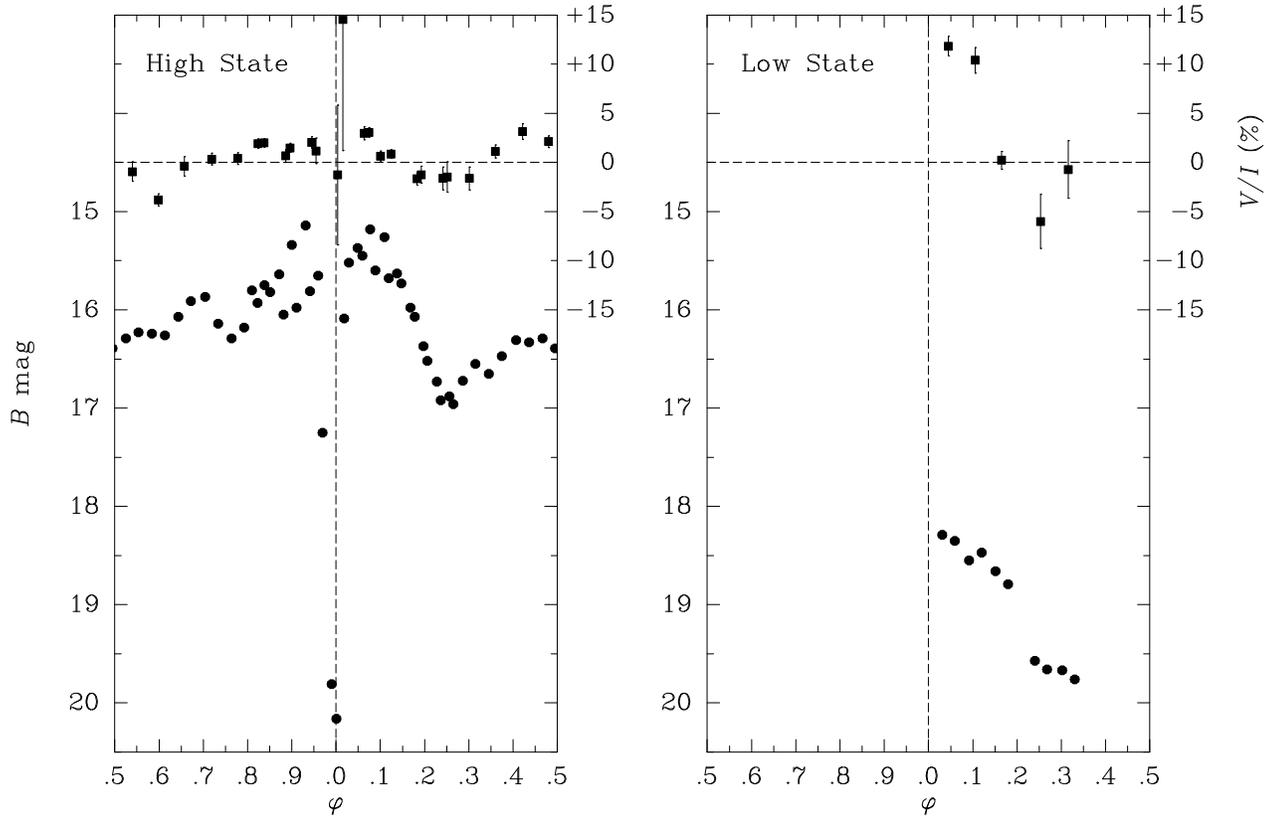}
\vspace{2.5truein}

\figcaption{{\it (Left:)\/} Spectrum-added circular polarization in per cent
{\it (squares)\/} and $B$-band brightness {\it (circles)\/} extracted from 1.5
orbits of spectropolarimetry obtained during a high state of SDSS~J0155+0028 in
2002 Dec.  The data are phase-folded on the ephemeris derived in $\S$4.1. Note
the weak polarization, which varies in sign through the cycle. {\it
(Right:)\/}  Results from a partial orbit acquired in a low accretion state in
2004 Feb.  Circular polarization in the low state is large during the interval
that the accretion spot is in view.}

\end{figure}

\clearpage

\begin{figure}
\includegraphics{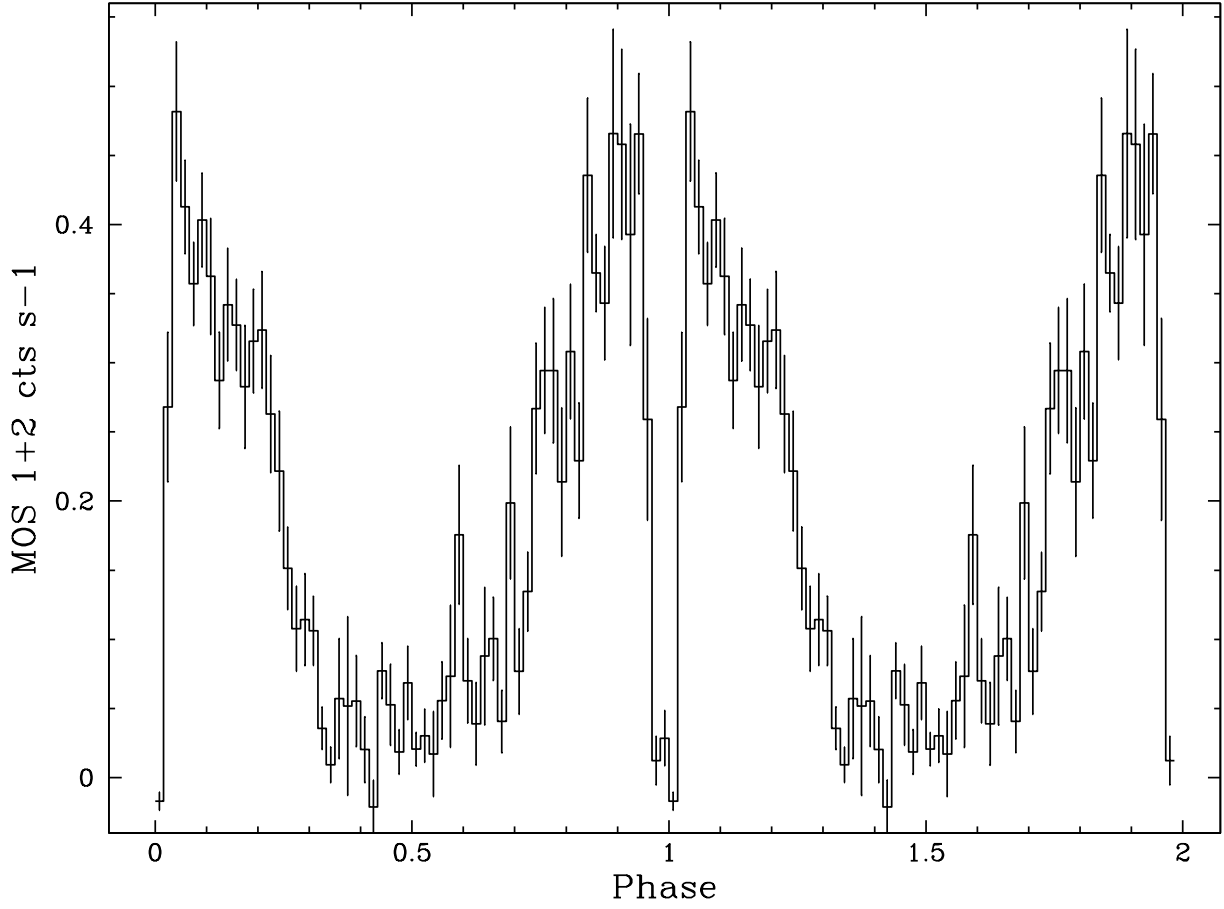}
\vspace{4.truein}

\figcaption{Mean light curve from the two {\it XMM-Newton}/EPIC-MOS cameras,
folded using the ephemeris derived in $\S$4.1.}

\end{figure}

\clearpage

\begin{figure}
\includegraphics{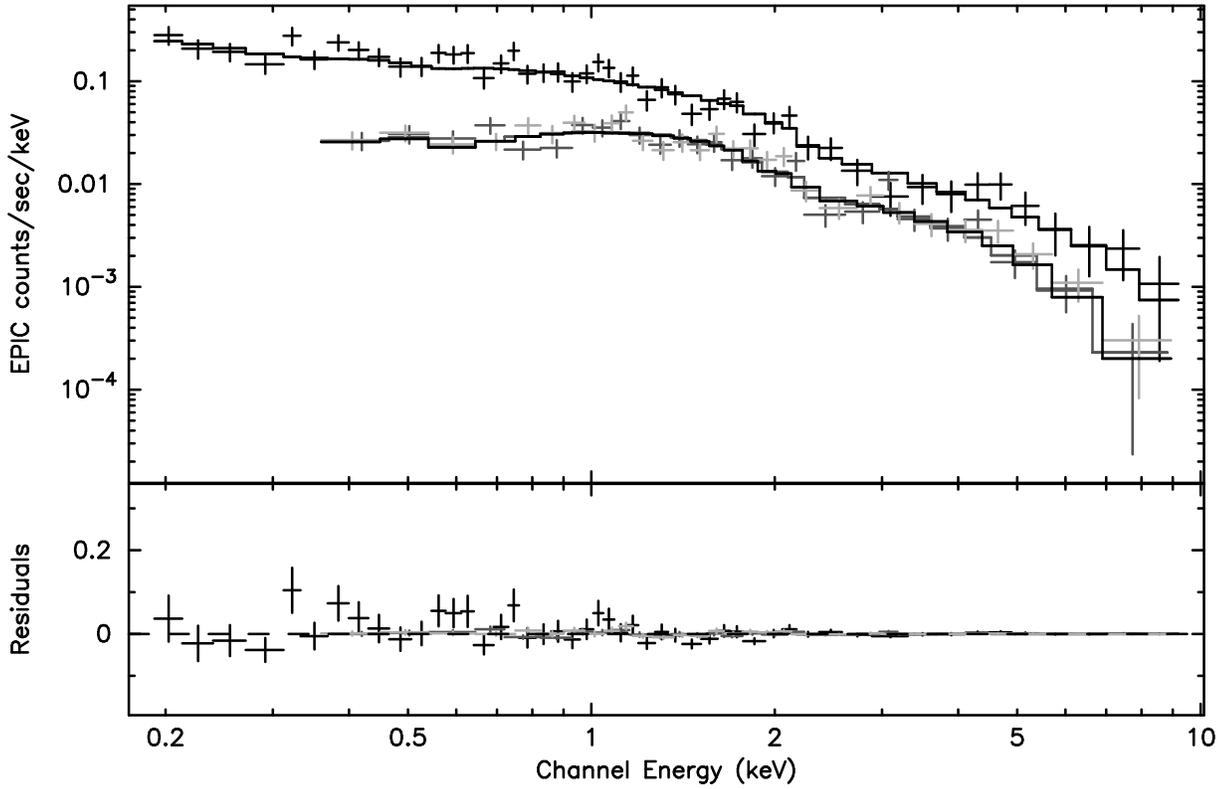}
\vspace{4.truein}

\figcaption{{\it XMM-Newton}/EPIC-pn {\it (upper)\/} and MOS {\it (lower)\/}
spectra.  The fit shown uses a 40~eV blackbody plus a 7~keV thermal
bremsstrahlung component (overplotted lines), although the inclusion of the
blackbody is not required at high significance.  Residuals to the fit are shown
in the lower panel.}

\end{figure}

\clearpage

\begin{figure}
\includegraphics{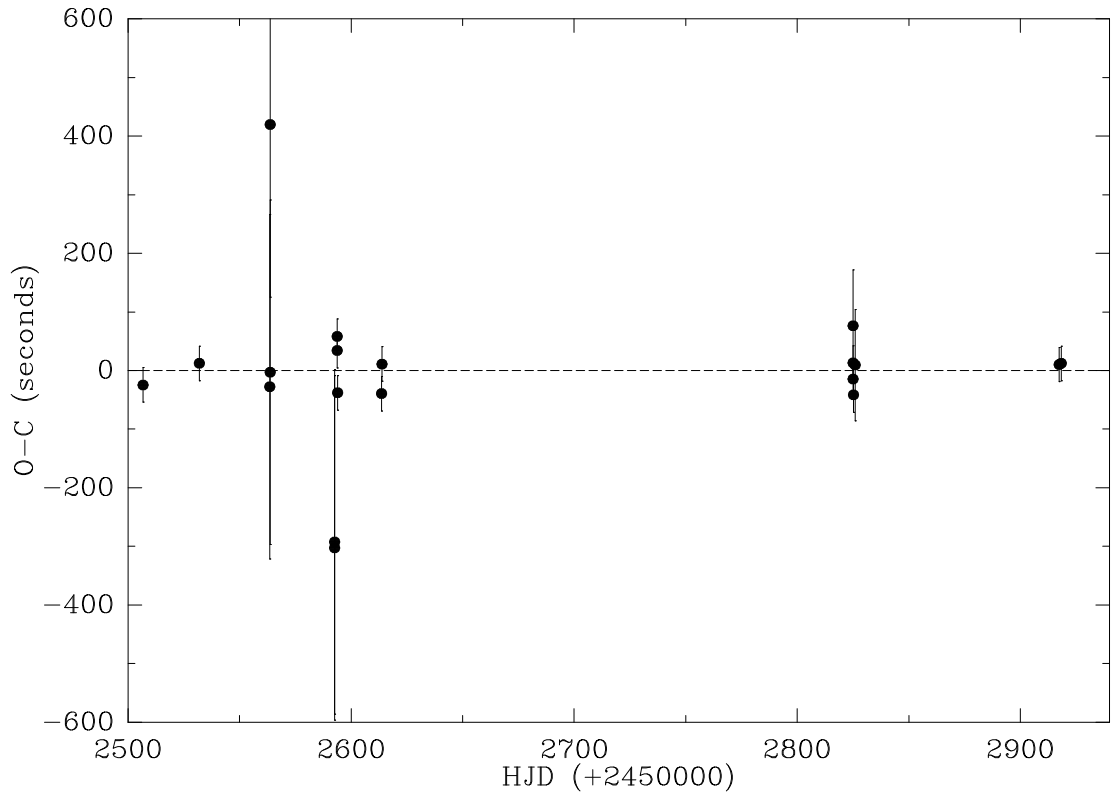}
\vspace{2.5truein}

\figcaption{An ``observed minus computed'' diagram for all 19 eclipse timings of
SDSS~J0155+0028 (Table 2), using the ephemeris derived in $\S$4.1.}

\end{figure}

\clearpage

\begin{figure}
\includegraphics{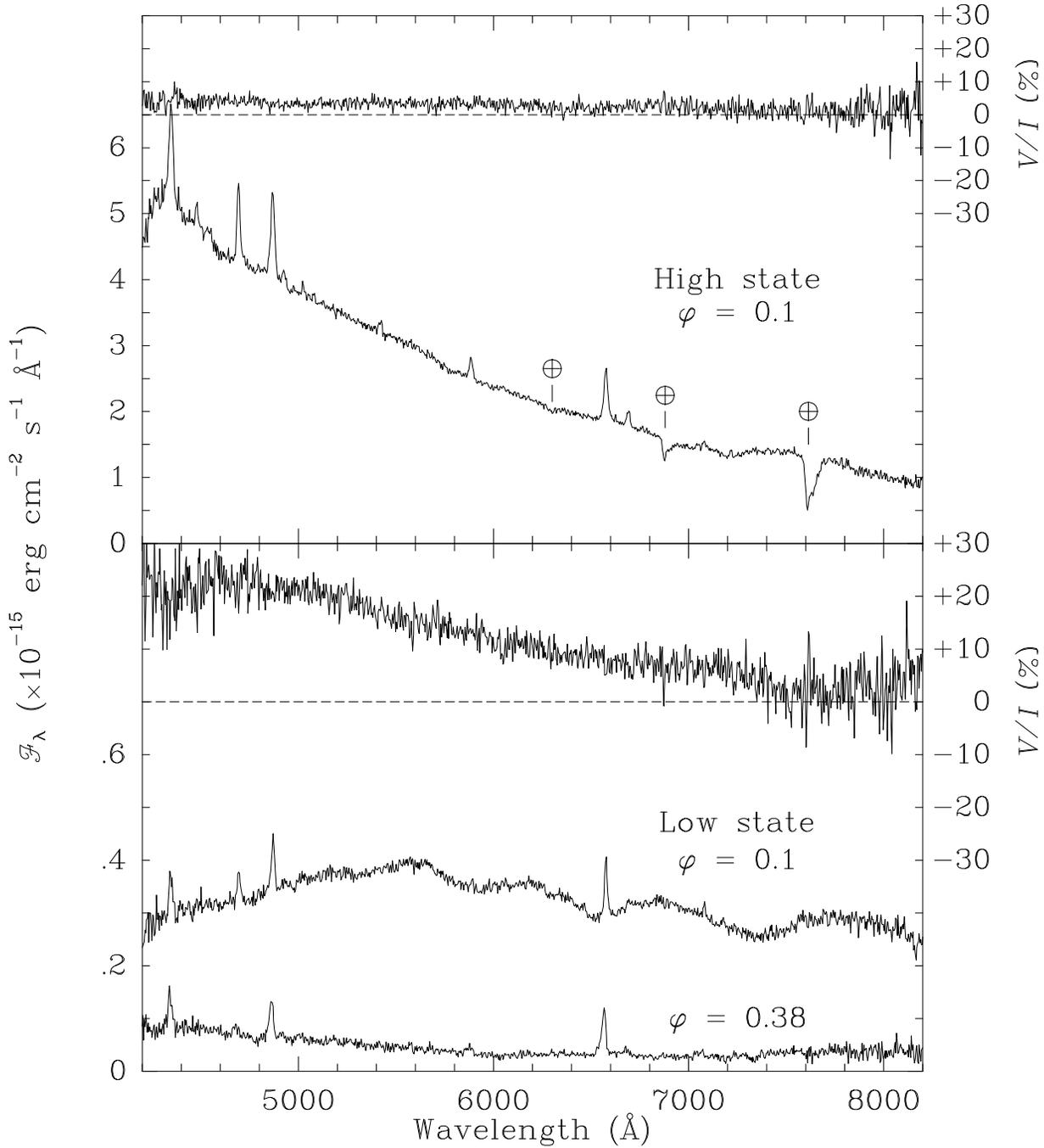}
\vspace{7truein}

\figcaption{Comparison of total flux and circular polarization spectra of
SDSS~J0155+0028 for high ($m_V=15.2$) and low accretion states ($m_V=17.5$),
both obtained soon after eclipse at $\varphi=0.1$.  The steep blue continuum,
inverted Balmer decrement, and smeared cyclotron harmonics in the high state
are indicative of a high-temperature shock, while the reduced polarization
throughout the spectrum results at least in part from the increased cyclotron
optical depth.   Note also the more widely-separated harmonics in the high
state; an indication of a second active pole.  The additional low-state
spectrum shown in the bottom panel represents the system when the active pole
is self-eclipsed by the white dwarf, and shows a slight upturn in the red due
to the cool companion star. Uncorrected terrestrial absorption in the upper
panel are marked.}

\end{figure}

\begin{figure}
\includegraphics{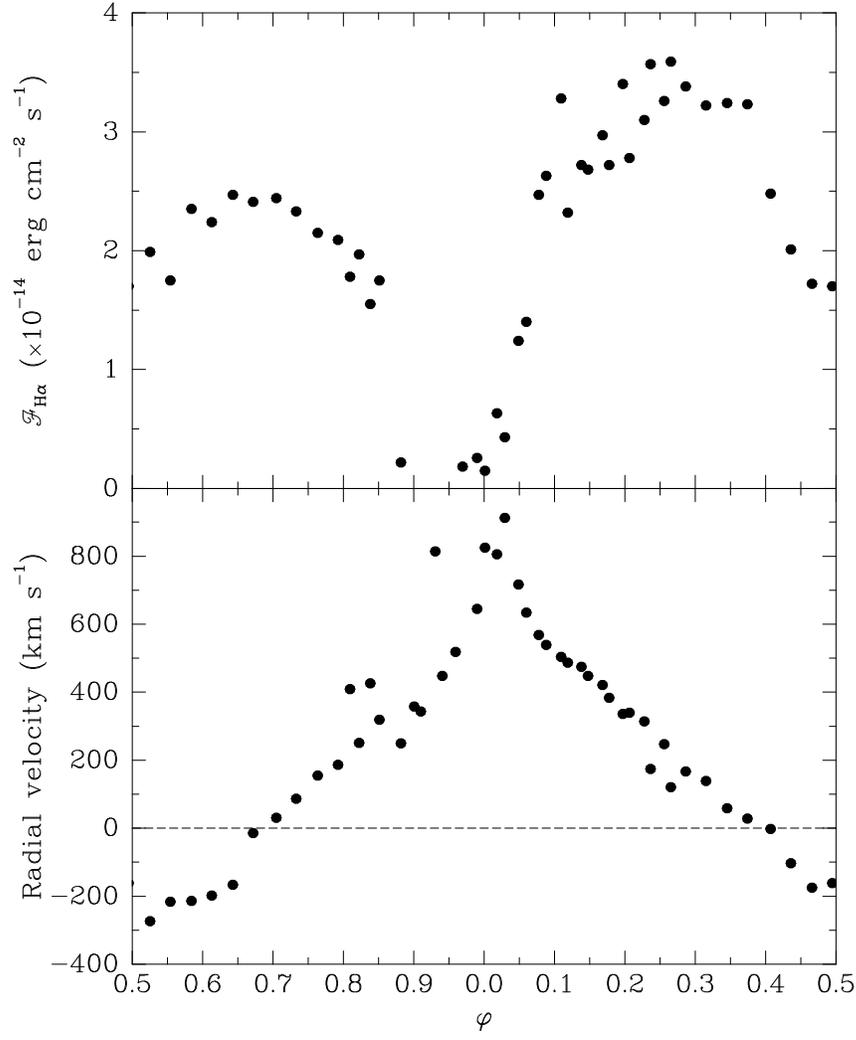}
\vspace{3.75truein}

\figcaption{Radial velocity curve {\it (bottom)\/} and line flux {\it (top)\/}
in H$\alpha$ for the high state of 2002 Dec.  Phases where the emission line
is absent or contaminated by absorption have been omitted.}

\end{figure}

\clearpage

\begin{figure}
\includegraphics{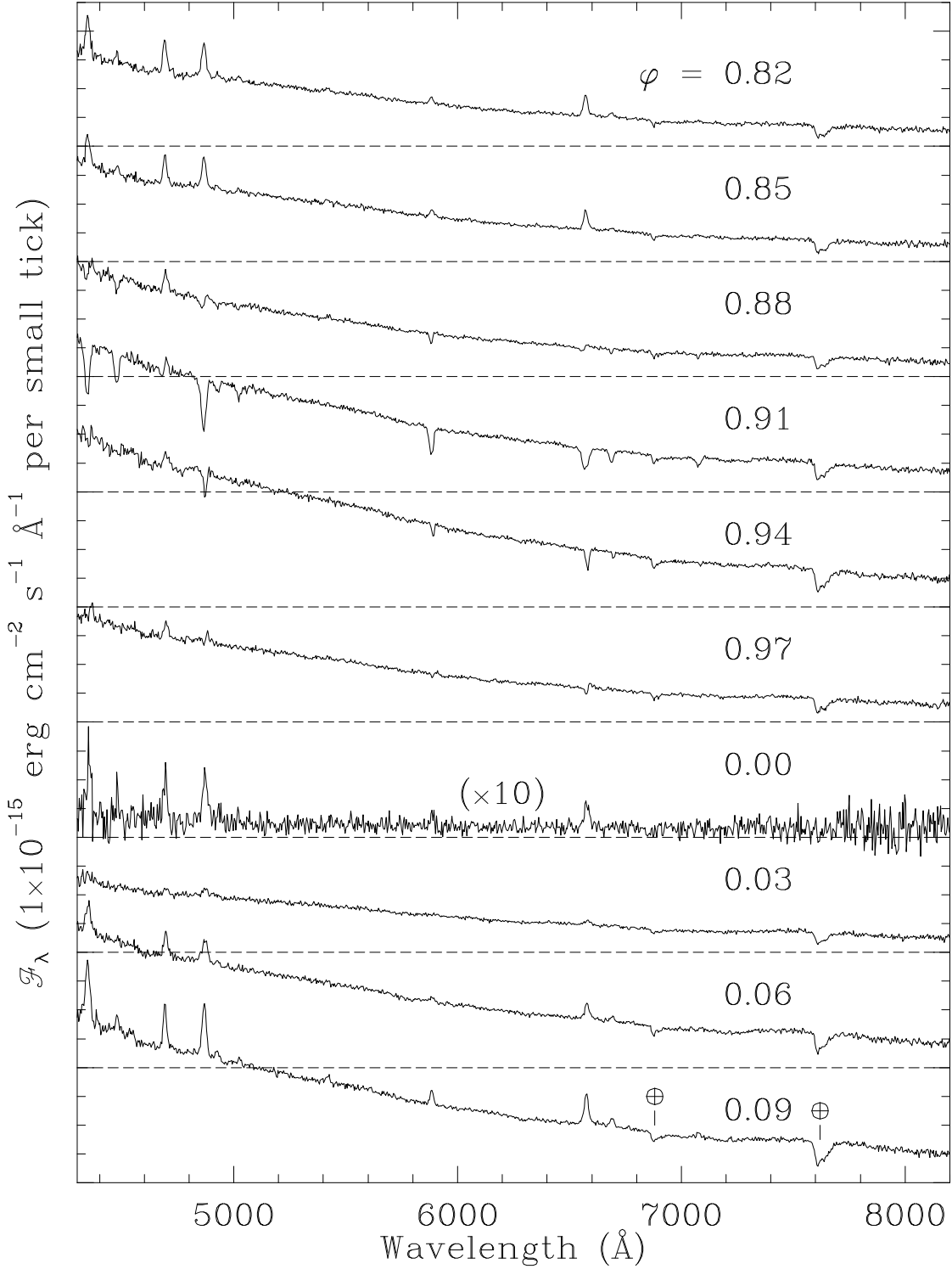}
\vspace{7.75truein}

\figcaption{Spectral sequence through eclipse during the high state of
SDSS~J0155+0028.  Time proceeds from top to bottom, with orbital phase
indicated.  Self-eclipse of the line-emitting stream begins at $\varphi=0.88$
(40$^\circ$ prior to primary eclipse) with the development of weak P-Cygni
profiles in the hydrogen Balmer and \ion{He}{1} lines.  This spectrum recurs
just prior to ingress of the primary eclipse.  Note that \ion{He}{2}
$\lambda$4686 never dips completely into absorption.  Dashed lines indicate
the zero-flux level of successive spectra.}

\end{figure}

\begin{figure}
\includegraphics{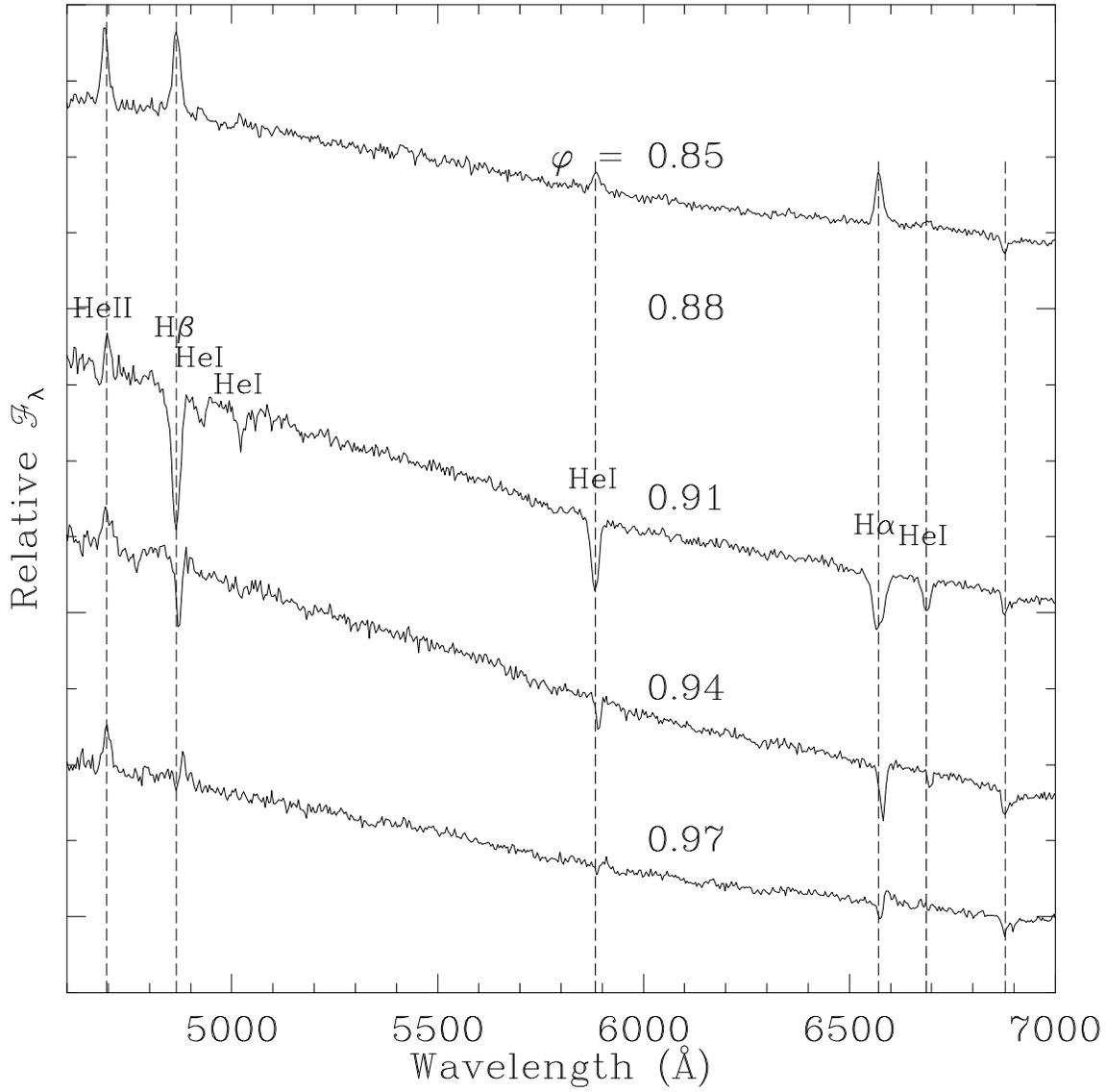}
\vspace{4.45truein}

\figcaption{Enlarged view of the spectral sequence leading up to primary
eclipse.  Note the development of P-Cygni features, followed by deep
absorption components for all lines except \ion{He}{2} $\lambda$4686.  The
absorption lines, which reach 50\% in depth for H$\beta$, become sharper
and shift to the red as primary eclipse approaches. The stationary feature at
6884~\AA\ is a terrestrial absorption bandhead of O$_2$.}

\end{figure}

\clearpage

\begin{figure}
\includegraphics{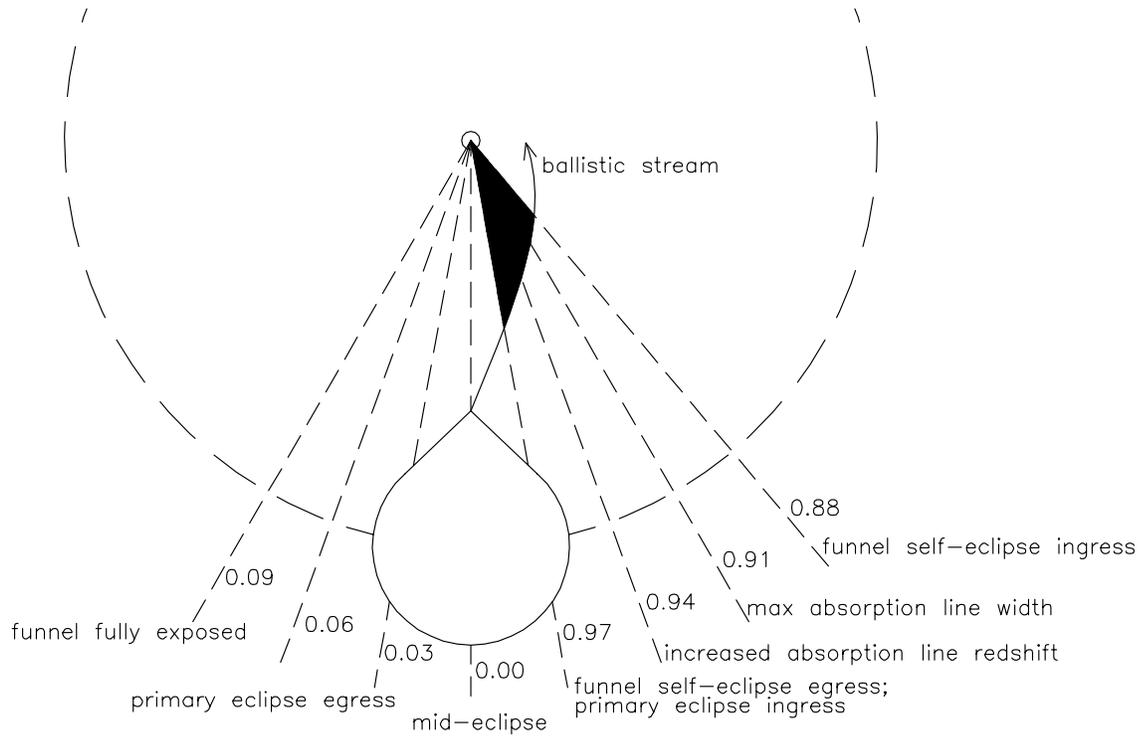}
\vspace{2.truein}

\figcaption{Sketch of SDSS~J0155+0028, shown to scale for an assumed
0.6~$M_\sun$ white dwarf and Roche-lobe filling, main-sequence secondary
($M_2=0.11~M_\sun$). The Earth lies very nearly in the orbital plane, with the
observer orbiting the binary clockwise with phase.  Also indicated is the
ballistic stream trajectory projected onto the orbital plane and viewing
phases corresponding to the spectra displayed in Figures 10 and 11.}

\end{figure}

\end{document}